\documentstyle[prl,aps,psfig,floats,array]{revtex}

\begin{document}

\wideabs{
\title{Antiferromagnetically coupled alternating spin chains
}
\author{Adolfo E. Trumper and Claudio Gazza }
\address{Instituto de F\'{\i}sica Rosario (CONICET) and
Universidad Nacional de Rosario, \\  Boulevard. 27 de febrero 210 bis,
(2000) Rosario, Argentina.}

\date{\today}
\maketitle
\begin{abstract}
The effect of antiferromagnetic interchain coupling in alternating
spin (1,1/2) chains is studied by mean of a spin wave theory and density 
matrix renormalization group (DMRG). In particular, two 
limiting cases are investigated, the two-leg ladder and its two dimensional (2D) 
generalization. Results of the ground state properties like energy, spin gap, 
magnetizations, and correlation functions are reported for the whole range of 
the interchain coupling $J_{\perp}$. For the 2D case the spin wave results
predict a smooth dimensional crossover from 1D to 2D keeping the
ground state  always ordered. For the ladder system, the DMRG
results show that any $J_{\perp}>0$ drives the system to a gapped
ground state.  Furthermore the behaviour of the correlation 
functions closely resemble the uniform spin-1/2 ladder. For $J_{\perp}$ 
lower than 0.3, however, the gap behaves quadratically as 
$\Delta\sim0.6 J^2_{\perp}$. Finally, it is argued that the behaviour of the 
spin gap for an arbitrary number of mixed coupled spin chains is analogous to
that of the  uniform spin-1/2 chains.\\

\end{abstract}
} 

\section{Introduction}
\begin{figure}[h]
\centerline{\psfig {figure=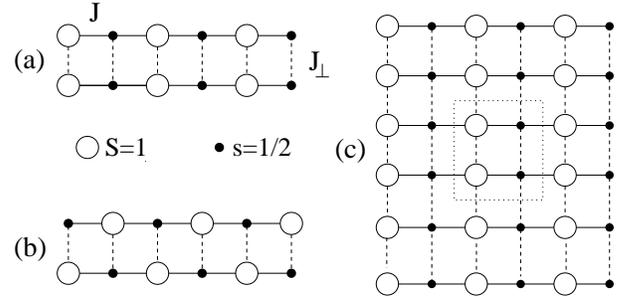,width=80mm,angle=0}}
\vskip 0.3cm
\caption{Three kind of alternating spin systems: (a) and (c) have a 
singlet ground state, (b) has a ferrimagnetic one.}
\label{2d}
\end{figure}

The recent synthesization of quasi-onedimensional bimetallic
magnets\cite{verdaguer,hagiwara}, with each unit cell containing two kind of 
different
spins,  has spurred a growing interest in the low-temperature properties of
quantum ferrimagnetic chains.
After intensive studies\cite{malvezzi,pati,tian,ivanov,yamamoto,yamamoto1,wu},
it was possible to describe properly the thermodynamic properties observed in
experiments by including the ferro and antiferromagnetic (AF)
features that  make this systems  specially attractive.
These results together with the possible experimental realization of new
mixed spin compounds motivated the study of a
variety of AF systems with different spin composition, all kind of
interactions, and topologies\cite{kawakami,kawakami1,fukui,satou}. From 
the experimental point of view, the ferrimagnetic compounds found are all 
composed of weakly coupled alternating spin chains\cite{verdaguer} 
(Fig.\ref{2d}(b) represents the ladder case
of such a topology). In other family of compounds like 
$MnCu(pba)(H_2O)_3.2H_2O$, with pba=1,3-propylenebis(oxamato), the ground 
state is a non-magnetic one, although it is composed of weakly coupled 
alternating spin ($Mn^{II}Cu^{II}$) chains\cite{verdaguer}
(Fig.\ref{2d}(a) and (c)). In fact, using the Heisenberg
Hamiltonian for these interacting
spin systems and applying Lieb-Mattis theorem\cite{mattis}, it is easy
to prove that in case (a) and (c) the ground state is AF with total spin
$S_{tot}=0$ whereas in case (b) it is ferrimagnetic with $S_{tot}\neq0$.
Nonetheless, in the 2D case we will see that the rotational spin symmetry can
be broken in the thermodynamic limit.
The aforementioned considerations motivated us to focus on the ground state
properties of an alternating spin ladder of
type (a) and a 2D array of type (c) with the following Hamiltonian,
\begin{equation}
\label{hamil}
H=J\sum_{<i,j>_{\|}} {\bf S_i}. {\bf s_j} +J_{\perp}
\sum_{<i,j>_{\perp}} \left( {\bf S_i}. {\bf S_j } + {\bf s_i}.
{\bf s_j}\right),
\end{equation}
where $S_i (s_i)$ represents a spin-1 (spin-1/2)and $<i,j>_{\|}$
($<i,j>_{\perp}$)
denotes nearest neighbors along horizontal (vertical) direction.
To our knowledge, this is the first systematic study of
antiferromagnetically coupled alternating spin chains.
Only recently, this issue has been addressed but for ferrimagnetic
ladders (Fig.\ref{2d}(b))\cite{langari,ivanovL}. Instead, in the alternating 
spin ladder we are interested in, (in what follows ASL), the global properties 
of the system change completely once the chains are coupled. Unlike uniform 
spin ladders, little is known about the effect of AF interchain coupling in ASL.
For instance, in a uniform spin-$1/2$ ladder (USL), it is well
known that an infinitesimal $J_{\perp}$ is required to send the
chains off-criticality --with an excitation gap growing linearly
with $J_{\perp}$-- independently of its
sign\cite{barnes,elbiorice}. Besides, it has been pointed out 
that the absence of broken symmetry in the chains is indeed the
crucial feature for such a  behaviour\cite{barnes}. On the other
hand, in the spin-$1$ case, it has been argued that a interchain 
coupling $J_{\perp}$ makes the gap (already existing in the
chains) to decrease with $J_{\perp}$, and again no matter the sign
of $J_{\perp}$\cite{senechal}. For the mixed case, we are
concerned, the situation is rather different as  the ferrimagnetic
chains are neither gapped nor critical, but they are gapless as a
result of a broken symmetry ground state. Hence, one could
intuitively say that the effect of interchain coupling should be
different to the both uniform cases mentioned above and a finite
amount of $J_{\perp}$ could be needed to melt order. We shall see,
however, that the low dimensionality of the problem  (1D) is
crucial and that any $J_{\perp}>0$  drives the system to a gapped
ground state. The study of the spin correlation functions confirms
this picture, showing a strong resemblance to USL. Contrary to the
ASL, in the 2D case we shall see that the dimensional crossover
from 1D is completely smooth and keeps  the ground state ordered
for the whole range of $J_{\perp}$.

The work is organized as follows: in section II we study
the dimensional crossover from 1D to 2D within spin wave theory. In
section III we give some details of the DMRG method implemented and 
discuss the results for the ladder case. In section IV we summarize 
the results and give the concluding remarks.

\section{Spin wave theory}
Though approximate, it has been shown that the
spin wave (SW) series converges correctly to the DMRG results in ferrimagnetic
chains\cite{ivanov}.
Hence SW appears to be, in principle, a good starting point technique to study
the crossover from 1D to 2D where SW is even more reliable. It is also known,
however, that for a ladder system SW breaks down\cite{fukui} (see below).
The problem requires a unit cell composed of two spins $S=1$ and two spins
$S=1/2$ (see Fig.\ref{2d}(c)). At the classical level, it is assumed a N\'eel
order and quantum fluctuations are incorporated by four kind of bosons leading
to the following  Holstein-Primakov transformation,
$$
\begin{array}{c}
S^z_A=-s_1+a^{\dagger} a \;\;\; S^+_A= a^{\dagger}\left(\sqrt{2s_1-a^{\dagger} a}\right)
\\

S^z_B  =  s_2-b^{\dagger} b\;\;\;  S^+_B  = \left(\sqrt{2s_2-b^{\dagger}b}\right) b \\

S^z_C=s_1-c^{\dagger} c\;\;\;   S^+_C= \left(\sqrt{2s_1-c^{\dagger} c}\right)c \\

S^z_D=-s_2+d^{\dagger} d \;\;\; S^+_D= d^{\dagger}\left(\sqrt{2s_2-d^{\dagger} d}\right).\\
\end{array}
$$
\noindent Introducing this representation  in
the Hamiltonian (\ref{hamil}), and Fourier transforming to  {\bf k}-space, 
the SW Hamiltonian can be written as,
$$
H_{sw}= const+\sum_k {\overline a}^{\dagger}_{\bf k}\left(\begin{array}{cc}
                                    d & \Delta \\
                                    \Delta & d
                                 \end{array}\right) {\overline a}_{\bf k}
$$
\noindent where ${\overline a}^{\dagger}_{\bf k}=(a^{\dagger}_{\bf k},
b^{\dagger}_{\bf -k},c^{\dagger}_{\bf -k},d^{\dagger}_{\bf k},a_{\bf k},b_{\bf -k},
c_{\bf -k},d_{\bf k})$, $d=\frac{1}{2}diag(A_1,A_2,A_1,A_2)$,

$$
\Delta=\frac{1}{2}\left(\begin{array}{cccc}
                                    0 & C_{\bf k} & D^{\perp}_{1{\bf k}} & 0  \\
                                    C_{\bf k} & 0 & 0 & D^{\perp}_{2{\bf k}}  \\
                                    D^{\perp}_{1{\bf k}} & 0 & 0 & C_{\bf k}   \\
                                    0 & D^{\perp}_{2{\bf k}} & C_{\bf k} & 0
                                   \end{array}\right),
$$
\noindent and
$$
\begin{array}{c}
 const=-4NJs1s2-N2J_{\perp}(s^2_1+s^2_2) \\

A_1=(2Js_2+2J_{\perp}s_1),\;\;\; A_2=(2Js_1+2J_{\perp}s_2) \\

D^{\perp}_{1{\bf k}}=2s1J_{\perp}\cos{ky/2}\;\;\;, D^{\perp}_{2{\bf k}}=2s2J_{\perp}\cos{ky/2}, \\

C_{\bf k}=2J\sqrt{s_1s_2}\cos{kx/2}. \\
\end{array}
$$
\begin{figure}
\centerline{\psfig {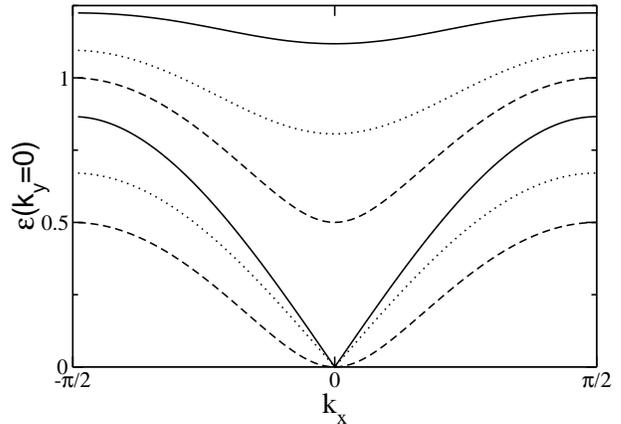}}
\caption{Spin wave spectrum for the 2D array. Dashed, dotted, and solid lines
are for $J_{\perp}=0, 0.2$, and $0.5$, respectively. For the ladder case the
spectrum is qualitatively the same.}
\label{spect}
\end{figure}
\noindent We have para-unitary diagonalized\cite{colpa} $H_{sw}$ which allowed
us to calculate the spectrum excitation (with an acoustic and an optic band 
each one twofold degenerate), the ground state energy and the magnetizations. 
For the particular case $J_{\perp}=0$ we recover the SW 
predictions\cite{pati,ivanov,yamamoto} of a ferrimagnetic chain, that is, a 
quadratic gapless dispersion near ${\bf k}=0$ (Fig.\ref{spect}) (which 
manifests the ferromagnetic character of the system at low-temperature), an 
antiferromagnetic gapped mode, and the magnetization values $m_1= 0.6951$
and $m_2=0.1951$ for the spin-1 and spin-1/2, respectively. As soon as
$J_{\perp}$ is switched on the gapless mode becomes linear reflecting the
antiferromagnetic character\cite{nota} of the coupled system,
whereas the optical modes moves upward (Fig.\ref{spect}). 
\begin{figure}[h]
\centerline{\psfig {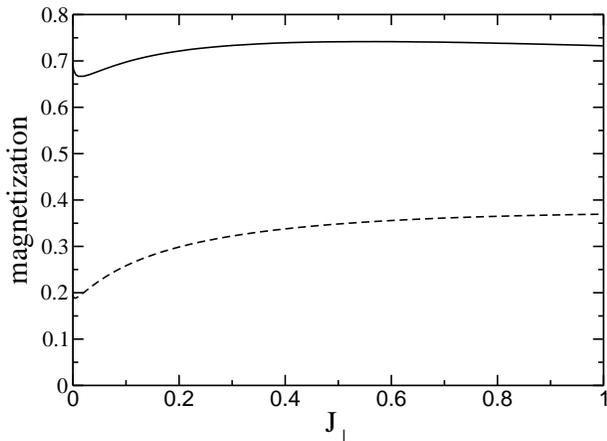}}
\caption{Magnetizations predicted by spin wave theory. Solid (dashed) line
represents the magnetization $m_1$ ($m_2$) for spin-1 (spin-1/2) sites.}
\label{magne}
\end{figure}
For the isotropic case $J_{\perp}=1$ we obtained a ground state energy 
$e_0=-1.3631$, and $m_1=0.7326$, $m_2=0.3694$ for the magnetizations. Notice 
that $m_2$ is bigger than the SW value of the uniform spin-1/2 case $m=0.303$.
Using mean-field
arguments, such a robust spin-1/2 magnetization is due to the more stronger
field caused by  the spin-1 neighbors (a similar reasoning explains why $m_1$
is smaller than the SW value of the uniform spin-1 case $m=0.803$).
It is also interesting to note that the relative reduction of the 
magnetizations, $m_1/S$ and $m_2/s$, are both nearly 0.73 in the isotropic 
case, while $m_1/S\sim0.70$ and $m_2/S\sim0.40$ for the ferrimagnetic chain.
In Fig.\ref{magne} we show $m_1$ and $m_2$ versus $J_{\perp}$. It can be seen
that the effect of 2D coupling is to enhance the values of the magnetizations,
being the crossover completely smooth.
On the other hand, our SW calculation can be adapted to the ladder case (a)
by just changing:
$$
\begin{array}{c}
 const=-4NJs1s2-NJ_{\perp}(s^2_1+s^2_2) \\
A_1=(2Js_2+J_{\perp}s_1),\;\;\; A_2=(2Js_1+J_{\perp}s_2) \\
D^{\perp}_{1{\bf k}}=s1J_{\perp}\;\;\;, D^{\perp}_{2{\bf k}}=s2J_{\perp}, \\
                  C_{\bf k}=2J\sqrt{s_1s_2}\cos{kx/2}. \\
\end{array}
$$ and ${\bf k}$ running along the 1D Brillouin zone. Now, the
linear behaviour of the gapless mode leads to important quantum
fluctuations that prevent the ordering in 1D. This feature was
previously pointed out by Fukui et. al \cite{fukui} and suggests a
possible opening of a spin gap which is impossible to predict
within a SW calculation. 
In spite of such break down we have
corroborated that the  SW ground state energy compare quite well
with the DMRG prediction.

\section{Density matrix renormalization group}
\begin{figure}[ht]
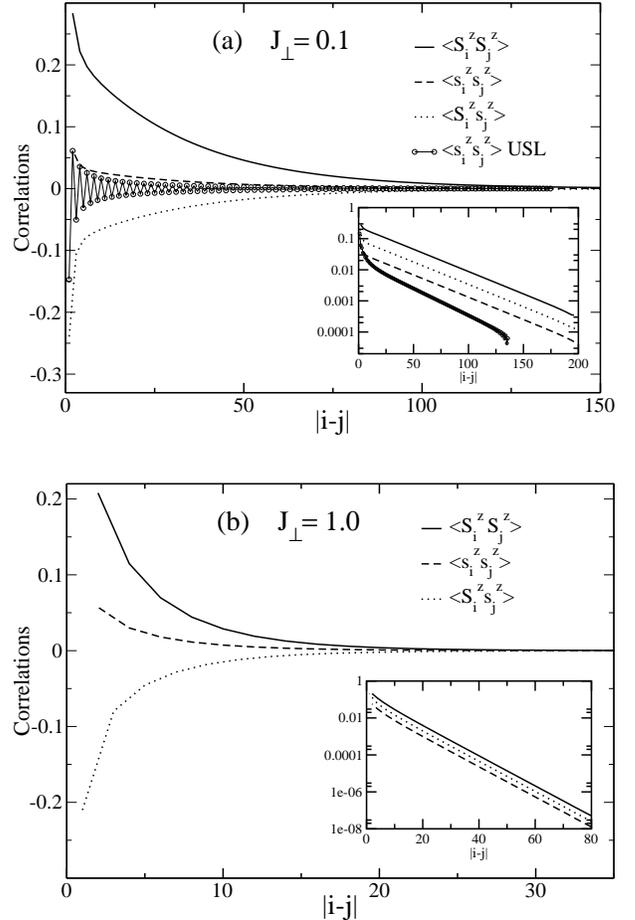

\centerline{\psfig {figure=correl01.eps,width=80mm,angle=0}}
\vskip 0.6 cm
\centerline{\psfig{figure=corrj1.eps,width=80mm,angle=0}}
\caption{Correlation functions obtained with infinite DMRG
algorithm, for (a) $J_{\perp}=0.1$ and (b) $J_{\perp}=1$. Inset:
semilog plot of the absolute values. In (a) the
results are compared with the USL.} \label{correl}
\end{figure}

The difficulty of SW theory to describe ASL leads us to implement
the DMRG method, which has shown a great versatility in different
quasi-onedimensional systems. We have performed calculations using
both, the finite and infinite algorithm, with open boundary
conditions (OBC). The procedure follows the usual DMRG
steps\cite{whiteorig}, thinking the ladder like a chain of rungs.
We found out that to avoid great edge effects, because of the OBC,
the definition of two different {\it single sites} is needed, each one
composed of a spin-1/2 rung and spin-1 rung, respectively (see
Fig.\ref{2d}(a)). The 4 and 9 states needed to span each rung
(otherwise we would need just 6 states) appear at the beginning as
a complication to the number of dominant density matrix states
kept, $m$, but they shown to be just of minor annoyance. Most of the 
calculations was carried out using $m$ between 200 and 400, getting 
a truncation error $O(10^{-8})$ at worst and $O(10^{-12})$ in the 
best case, reassuring us the reliability of the calculation. We have
calculated the triplet spin gap which is defined as 
$\Delta(L)=E(L,S_z=1)-E(L,S_z=0)$, with $E(L,S_{z})$ the ground
state energy for a {\it chain} with $L$ rungs ($2\times L$ is the
number of sites), and $z$ component of total spin 
$S_{z}$\cite{AFbehaviour}. For each $J_{\perp}$, we have 
extrapolated the gap with $L$ ranging from 10 to 100 rungs using a 
polynomial fit of the form\cite{white}:
\begin{equation}
\label{scaling}
\Delta(L)=\Delta+a_1/L+a_2/L^2+a_3/L^3+....
\end{equation}
\begin{figure}
\centerline{\psfig{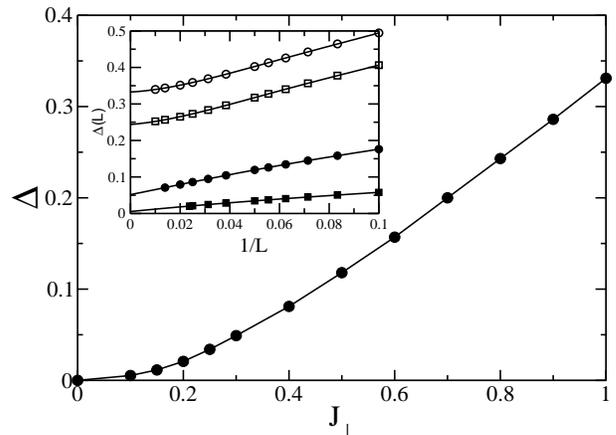}}
\caption{Spin-gap versus $J_{\perp}$ obtained with DMRG. Inset:
Spin gap as a function of $1/L$ for $J_{\perp} = 1$ (empty
circle), $0.8$ (empty square), $0.3$ (filled circle), and $0.1$
(filled square). Solid lines are polynomial fits as in
eq.(\ref{scaling}).} 
\label{gap}
\end{figure}
In Fig.\ref{gap} it is plotted the gap $\Delta$ vs. $J_{\perp}$.
In the particular case, $J_{\perp}=1$, we obtained a value for the
gap of $\Delta=0.334$.
For small values of $J_{\perp}$ (below 0.3)
the gap of the ASL behaves quadratically as $\Delta\sim 0.6
J^2_{\perp}$ and beyond $0.3$ it turns out quite linear 
corresponding to the strong coupling regime.
Consistently with these features, when extrapolating the
gap for different values of $J_{\perp}$, we detected a crossover
between two different scaling regimes around $J_{\perp}\sim 0.3$.
This is shown in the inset of Fig.\ref{gap} where we have plotted the 
scaling for $J_{\perp}=0.1, 0.3, 0.8$ and $1$. It is interesting to 
compare our result for the weakly coupled regime with that of the 
USL, which is known to behave linearly as 
$\Delta\sim 0.41J_{\perp}$\cite{greven}. 
\begin{figure}[h]
\centerline{\psfig {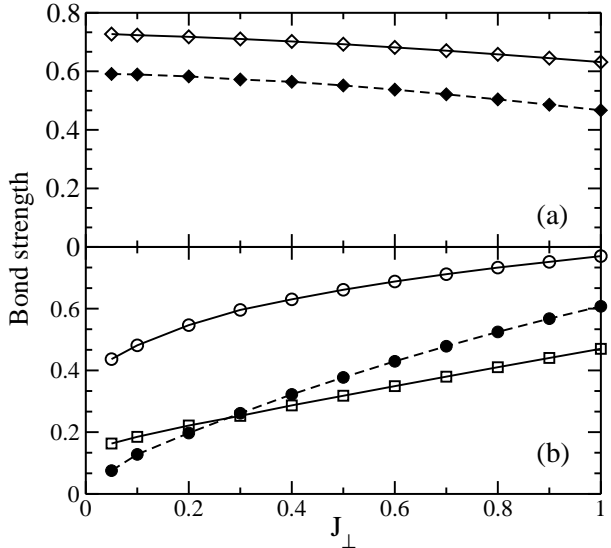}}
\caption{Normalized bond strength as a function of $J_{\perp}$.
(a) empty (filled) diamond along the leg for ASL (USL), and (b)
spin-1 rung (empty circle), spin-1/2 rung (empty square) for ASL and 
spin-1/2 rung for USL (filled circle).
} \label{vrcorrel}
\end{figure}
We think that this discrepancy reflects the quite distinct 
underlying physics behind the decoupled regime, that is, a critical 
state for uniform spin chains and a broken symmetry state for the 
ferrimagnetic chains.
In order to get an insight of the ground state configuration of
the ASL we computed\cite{nota1} the local bond strengths 
and compared them with the USL. The comparison was made after normalizing 
each value to its free bond case. Hence, for the very strong coupling regime
the values of the bonds along the leg (rung) must go to zero (one). 
Actually, we calculated the ${\it z}$ component of these quantities and 
because of rotational invariance we multiply it by a factor 3.
In Fig.\ref{vrcorrel}(a), we present the normalized values of  
$\langle {\bf S}_i.{\bf s}_{i+1} \rangle$ along the leg. 
A similar monotonic behavior 
with $J_{\perp}$ is observed for ASL and USL. On the other hand, in 
Fig.\ref{vrcorrel}(b) we show the normalized values of 
$\langle {\bf S}_i.{\bf S}_{i+1}\rangle$ and 
$\langle {\bf s}_i.{\bf s}_{i+1} \rangle$,
corresponding to the vertical spin-1 rungs and spin-1/2 rungs respectively.
It can be noticed that the spin-1 rungs are always larger than the spin-1/2
ones, however,  what is really remarkable is the crossing between the spin-1/2 
rungs values for ASL and USL which occurs just at $J_{\perp}\sim 0.3$. Even 
if we have not a microscopic explanation for that  we believe that  
our results for the gap and the local bond strengths are connected  and 
indicate a change of regime for the above value of $J_{\perp}$. 
For the sake of completeness, we studied the correlation functions along the 
legs. In Fig.\ref{correl}(a), we compare the spin correlations of
the USL with the three possible correlations of the mixed spin
case, corresponding to $J_{\perp}=0.1$\cite{nota2}. In the inset it is shown in
a semilog plot the exponential behaviour of the correlations with
an estimated correlation length of $\xi^{USL}\sim 25$, and $
\xi^{ASL}\sim 30$ lattice spacing. It can be noticed, also, that
the three possible correlation lengths are the same. We have checked that
the correlation functions along the leg decay exponentially for
the whole range of interchain coupling and the correlation length
$\xi^{ASL}\rightarrow\infty$ in the limit $J_{\perp}\rightarrow 0$
like in  USL\cite{greven}. When $J_{\perp}$ is strictly
zero there is a transition to two ferrimagnetically ordered chains
in the ASL.
In Fig.\ref{correl}(b) we show the correlations for $J_{\perp}=1$.
Again, the semilog plot (inset) shows the exponential decaying of
the correlations with an estimated correlation length of
$\xi^{ASL}\sim5$. This value is bigger than the isotropic one 
$\xi^{USL}\sim 3$ found for the uniform case\cite{white,greven}. 

\section{Summary and concluding remarks}
In summary, we have studied the effect of AF coupling in mixed
spin chains for two particular topologies, the ladder and the two
dimensional system. In the latter, SW theory leads to a  smooth
dimensional crossover from 1D to 2D  with an  enhancement of the
magnetizations. Furthermore, as a consequence of the crossover, we 
detected an important difference
in the relative magnetizations, being $m_1/S$ and $m_2/s$ around
0.73 for $J_{\perp}=1$ whereas $m_1/S\sim0.70$ and $m_2/s\sim0.40$
for $J_{\perp}=0$. For the ladder system, where SW breaks down, we
used DMRG to study the gap and the correlation functions with
$J_{\perp}$. For $J_{\perp}\neq 0$ the ladder system is always
gapped with  a quadratic behaviour 
$\Delta\sim0.6 J^2_{\perp}$ for $J_{\perp}$ lower than $0.3$, and 
then it becomes quite linear for greater values of the interchain coupling. 
The study of the local bond strengths also indicate a change of regime 
at $J_{\perp}\sim0.3$. In the particular case of $J_{\perp}=1$ we found that the 
gap is $\Delta=0.334$ and that the correlation functions  decay exponentially   
with a correlation length about $\xi^{ASL}\sim5$. 
Except for the differences found in the weakly coupled regime, our results 
suggest a strong similarity between the ASL and the USL. 
This close resemblance encourage us to
make some general statement regarding an arbitrary number of mixed
spin chains, between  the two limiting cases we have studied.
First, we can rigorously say that any odd number of chains has
always a ferrimagnetic ground state, due to Lieb-Mattis theorem,
so, it will be gapless --but ordered--. Then, if we complement our
results with the conjecture that an even number of chains will be
gapped and --similarly to uniform
ladders\cite{elbiorice,white,rice}-- this gap will decrease to
zero in the 2D limit, it is recovered an analogous spin gap
behaviour to the uniform spin-1/2 case. In both cases, however,
the  nature of the gapless states  is completely different. 
We hope our findings could be tested experimentally in a near future.

\acknowledgments
The authors would like to thank J. Riera, A. Dobry and O. L. Manuel for 
very fruitful discussions. This work was partially supported by 
Fundaci\'on Antorchas.


\begin{references}

\bibitem{verdaguer} Y. Pei, M. Verdaguer, O. Kahm, J. Sletten and J. P. Renard, Inorg. Chem.
{\bf 26}, 138 (1987); O. Kahm, Y. Pei, M. Verdaguer, J. P. Renard and J. Sletten.
J. Am. Chem. Soc. {\bf 110}, 782 (1988).

\bibitem{hagiwara} M. Hagiwara, K. Minami, Y. Narumi, K. Tatani, and K. Kindo,
J. Phys. Soc. Jpn. {\bf 67}, 2209 (1998).

\bibitem{malvezzi}F. C. Alacraz and A. L. Malvezzi, J. Phys. Math. Gen. {\bf 30}, 767 (1997).

\bibitem{pati} S. K. Pati, S. Ramasesha and D. Sen, \prb {\bf 55}, 8894 (1997)

\bibitem{tian} G. S. Tian, \prb {\bf 56}, 5355 (1997).

\bibitem{ivanov} N. B. Ivanov \prb {\bf 57}, R14024 (1998); 
N. B. Ivanov, \prb {\bf 62}, 3271 (2000).

\bibitem{yamamoto}S. Yamamoto, S. Brehmer adn H. J. Mikeska, \prb {\bf 57}, 13610 (1998);
\prb {\bf 57}, R14008 (1998); S. Yamamoto, T. Fukui, K. Maisinger and 
U. Schollwock, Phys.
Condens. Matter {\bf 10}, 11033 (1998); S. Yamamoto, T. Fukui and T. Sakai, Eur. Phys.
J. B {\bf 15}, 211 (2000)
T. Fukui, \prb {\bf 57}, R14008 (1998).

\bibitem{yamamoto1}S. Yamamoto, S. Brehmer adn H. J. Mikeska, \prb {\bf 57}, 13610 (1998).

\bibitem{wu} C. Wu, B. C. Xi Dai, Y. Yu and Z. B. Su, \prb {\bf 60}, 1057 (1999).

\bibitem{kawakami}A. Koga, S. Kumada, N. Kawakami and T. Fukui, J. Phys. Soc. Jpn. {\bf 67},
622 (1998); T. Fukui and N. Kawakami, \prb {\bf 56}, 8799 (1997);
 A. Koga, S. and N. Kawakami, J. Phys. Soc. Jpn.
{\bf 69}, 1834 (2000).

\bibitem{kawakami1}Y. Takushima, A. Koga and N. Kawakami, \prb {\bf 61}, 6133 (2000).


\bibitem{fukui}  T. Fukui and N. Kawakami, \prb {\bf 57}, 398 (1998).

\bibitem{satou} A. Satou and Y. Nakamura, J. Phys. Soc. Jpn {\bf 68}, 4014 (1999).

\bibitem{mattis} E. Lieb and D. Mattis, J. Math. Phys.{\bf 3}, 749 (1962).

\bibitem{langari} A. Langari, M. Abolfath and M. A. Martin-Delgado, \prb {\bf61}, 343 (2000); A. Langari and M. A. Martin-Delgado, \prb {\bf 62}, 11725 (2000); A. Langari and M. A. Martin-Delgado, \prb {\bf 63},
54432 (2001).

\bibitem{ivanovL} N. B. Ivanov and J. Richter, Condmat/0011388.



\bibitem{barnes} T. Barnes, E. Dagotto, J. Riera and E. S. Swanson, \prb {\bf 47}, 3196 (1993).

\bibitem{elbiorice} E. Dagotto and T. M. Rice, Science {\bf 271}, 618 (1996).

\bibitem{senechal} D. Allen and D. S\'en\'echal, \prb {\bf 61}, 12134 (2000).

\bibitem{colpa}  J. H. P. Colpa, Physica {\bf 93A}, 327 (1978).


\bibitem{nota} This behaviour has been also reported in ref.\cite{kawakami1} using 
a Schwinger boson mean-field theory.


\bibitem{whiteorig} S. R. White \prl {\bf 69}, 2863 (1992); S. R. White \prb
{\bf 48}, 10345 (1993).

\bibitem{AFbehaviour} In order to check the  AF behaviour of the coupled
ladder we verified the equivalence between the spin gap excitations
$\Delta Sz=1$ and $\Delta Sz=-1$. This should be compared  with the
ferrimagnetic case where there is an asymmetry in the spin excitation:
one gapless for $\Delta Sz=-1$ and other gapped for $\Delta Sz=1$.

\bibitem{white} S. R. White, R. M. Noack and D. J. Scalapino. \prl {\bf 73}, 886 (1994).

\bibitem{greven} M. Greven, R. J. Birgeneau, and U. J. Wiese, \prl {\bf 77}, 1865 (1996).

\bibitem{nota1} For this calculation we used the infinite DMRG algorithm, and we 
carried out the measurements on the two central rungs inserted at the last 
iteration, in order to describe the bulk properties when convergence was reached.

\bibitem{nota2} In all cases the calculation of the spin correlation as a 
function of $|i-j|$ is performed with $i$ located at the center of the ladder 
and both sites belonging to the same leg. 

\bibitem{rice} T. M. Rice, S. Gopalan and M. Sigrist, Europhys. Lett. {\bf 23},
445 (1993); S. Gopalan, T. M. Rice and M. Sigrist, \prb {\bf 49}, 8901 (1994);
B. Frischmuth, B. Ammon and M. Troyer, \prb {\bf 54}, R3714 (1996).

\end{references}
\end{document}